\documentclass[]{emulateapj}
\begin{document}
\title{CN Bimodality at Low Metallicity: The Globular Cluster M53}

\bigskip
\author{Sarah L. Martell, \altaffilmark{1} Graeme H. Smith, \altaffilmark{1,2} and Michael M. Briley \altaffilmark{3}}
\altaffiltext{1}{Department of Astronomy \& Astrophysics, UC Santa Cruz, martell@ucolick.org}
\altaffiltext{2}{University of California Observatories/Lick Observatory, graeme@ucolick.org}
\altaffiltext{3}{Department of Physics and Astronomy, University of Wisconsin Oshkosh, mike@maxwell.phys.uwosh.edu}
\bigskip
\begin{abstract} 
\noindent

We present low resolution UV-blue spectroscopic observations of red giant stars in the globular cluster M53 ([Fe/H]$=-1.84$), obtained to study primordial abundance variations and deep mixing via the CN and CH absorption bands.  The metallicity of M53 makes it an attractive target: a bimodal distribution of 3883 \AA\ CN bandstrength is common in moderate- and high-metallicity globular clusters ([Fe/H] $\geq -1.6$) but unusual in those of lower metallicity ([Fe/H] $\leq -2.0$).  We find that M53 is an intermediate case, and has a broad but not strongly bimodal distribution of CN bandstrength, with CN and CH bandstrengths anticorrelated in the less-evolved stars.  Like many other globular clusters, M53 also exhibits a general decline in CH bandstrength and [C/Fe] abundance with rising luminosity on the red giant branch.  
\end{abstract}


\section{Introduction}
Star-to-star variations of the strength of the 3883 \AA\ CN band are nearly universal in moderate-metallicity ([Fe/H] $\geq -1.6$) globular clusters of the Milky Way, but unusual in low-metallicity globular clusters ([Fe/H] $\leq -2.0$) (see Gratton, Sneden \& Carretta 2004 for a thorough review).  Typically a CN bandstrength is measured as a magnitude difference between the flux within the 3883 \AA\ CN band and the flux in a nearby comparison band chosen to be free of CN absorption.  The index $S(3839)$ introduced by Norris et al. (1981) is one such measure of CN band strength.  Not only does $S(3839)$ vary from star to star within individual globular clusters, its distribution is often bimodal in moderate-metallicity ([Fe/H] $\geq -1.6$) clusters.  It is often anticorrelated with the strength of CH absorption in the G band at 4300 \AA\ among giants of similar absolute magnitude.  This anticorrelation indicates that CN-strong stars are depleted in carbon.  In order to have strong CN bands with low [C/Fe] abundances, the CN-strong stars must also be enhanced in nitrogen.  This abundance pattern has directed investigations into the origin of the bimodality toward stellar sources, where hydrogen shell burning can produce abundance enhancements and depletions such as those observed in globular clusters in C and N (and also O, Na, Mg, and Al).  Meanwhile, few CN-strong giants are known in the halo field (see, e.g., Pilachowski et al. 1996; Shetrone et al. 1999), indicating that the enrichment process responsible for CN bimodality only operated in cluster (or proto-cluster) environments.  This paper considers whether the apparent lack of CN bimodality in low-metallicity globular clusters is a true lack of abundance variation or simply an effect of low overall metal abundance.

Previous investigations into the behavior of CN abundance at low metallicity have been inconclusive as to whether large variations in bandstrength can be present.  Trefzger et al. (1983) and Shetrone et al. (1999) report several CN-strong red giants with large nitrogen abundances in M15 ([Fe/H]$=-2.26$), and Carbon et al. (1982) found variations of [C/Fe] and [N/Fe] but no strong 3883 \AA\ CN bands among the red giant stars of M92 ([Fe/H]$=-2.28$).  Briley et al. (1994) showed that the carbon and nitrogen abundances measured within M92 would result in a bimodal distribution of $S(3839)$ if the overall metallicity were scaled up to [Fe/H]$=-1.0$ dex.  Smith \& Norris (1982) and Briley et al. (1993) both studied M55 ([Fe/H]$=-1.81$, Harris 1996), and found that it is quite homogeneous in CN bandstrength.  That lack of CN-strong stars could be the result of low overall metallicity inhibiting the formation of CN molecules, low [C/Fe] among the brighter red giants, or a sign that whatever enrichment process caused CN variations in moderate-metallicity globular clusters did not operate at lower metallicities.  A study of additional globular clusters with metallicities similar to M55 could help decide between these alternatives.  We chose M53 as the subject of this study because it is similar in metallicity to M55 (based on our isochrone fits discussed below), and will hopefully provide insight into whether low-metallicity globular clusters contain CN-enhanced stars.  While our study does not comment directly on the process that causes abundance inhomogeneities in globular clusters, we note here that a lack of abundance variation in the lowest-metallicity clusters would provide a constraint on all hypotheses about that process.

To understand the CN and CH index behavior, we compare spectroscopy of the 3883 \AA\ CN band for red giants in M53 to the Norris et al. (1981) study of NGC 6752, a moderate-metallicity globular cluster ([Fe/H]$=-1.56$) that shows clear CN bimodality.  We also convert CH bandstrengths into [C/Fe] abundances using synthetic spectra and look at the relations between carbon abundance and CN bandstrengths.  An intrinsic spread is found in CN bandstrength, but it is smaller than the range observed in NGC 6752.

This result is interesting in the light of work by D'Antona et al. (2002), who use stellar evolution models to draw connections between helium abundance and horizontal branch morphology.  They find that globular clusters with solar helium abundances have short, red horizontal branches, while extended blue horizontal branches result from the smaller core mass of helium-enhanced stars.  In their model, the helium enrichment comes from AGB feedback, implying that globular clusters with red horizontal branches, like M53, should exhibit only minor CNO-product enrichment.

\begin{figure}
\plotone{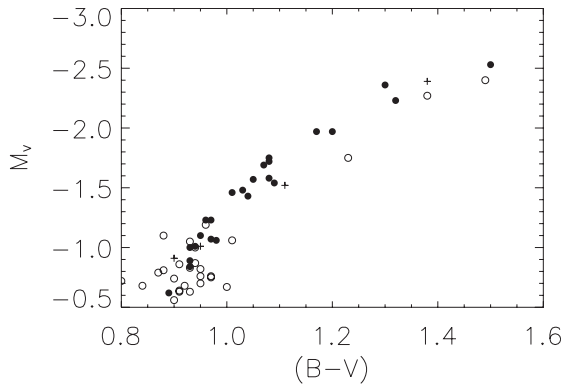}
\caption{
Color-magnitude diagram for the red giant branch of M53.  The photometry is from Rey et al. (1998) for stars identified as red giants in Cuffey (1965).  Filled circles are for stars observed with Kast, open circles are stars we did not observe, crosses are for stars dismissed in $\S$2 as nonmembers.}
\end{figure}

\section{The Data Set}

\begin{deluxetable*}{|l|r|r|l|c|c|c|c|c|c|c|}
\tablewidth{0pt}
\tablehead{\colhead{Star ID\tablenotemark{a}} & 
\colhead{X offset ($\arcsec$)\tablenotemark{b}} & \colhead{Y offset ($\arcsec$)\tablenotemark{b}} & \colhead{Date} & \colhead{$V$ \tablenotemark{b}} & \colhead{$(B-V)$ \tablenotemark{b}} & \colhead{t (s)} & \colhead{$S(3839)$} & \colhead{$S(CH)$} & \colhead{$CH(4300)$} & \colhead{[C/Fe]}}
\startdata
K&-139.7&-85.24&04152001&13.78&1.50&2400&0.1220&0.6081&0.5199&-0.35\\
S4 R4 16&56.59&-269.46&05162001&13.95&1.30&3600&0.0260&0.5693&0.4858&-0.65\\
G&-261.67&99.70&04152001&14.08&1.32&4200&0.0927&0.6385&0.5412&-0.35\\
S4 R4 24&-96.74&-252.57&05072002&14.34&1.20&3600&-0.0403&0.6639&0.6180&-0.35\\
S3 R2 22&-114.81&-4.25&05082002&14.34&1.17&1800&0.0961&0.5706&0.5580&-0.80\\
S4 R5 1&316.48&-281.80&05162001&14.56&1.08&3600&-0.0868&0.6693&0.6719&-0.40\\
M&-102.07&-163.22&04152001&14.59&1.08&2400&0.0198&0.6941&0.7028&-0.30\\
f&144.35&-285.43&05162001&14.62&1.07&3600&-0.0827&0.7090&0.6955&-0.20\\
Z&23.83&242.97&05082002&14.73&1.08&3600&0.0831&0.6680&0.6421&-0.40\\
N&6.19&-187.29&05072002&14.74&1.05&3600&-0.0664&0.7417&0.7363&-0.10\\
S3 R4 23&-278.38&151.07&04152001&14.77&1.09&2400&-0.0313&0.7194&0.7081&-0.20\\
P&104.23&-197.34&05162001&14.83&1.03&3600&-0.0758&0.6699&0.6379&-0.45\\
S4 R4 25&-104.09&-252.82&05072002&14.85&1.01&4500&-0.0292&0.6401&0.5963&-0.55\\
V&-109.99&210.18&05092002&14.88&1.04&3600&-0.1208&0.7124&0.7076&-0.25\\
S3 R2 3&-77.70&-105.43&05082002&15.08&0.97&3600&-0.1037&0.6821&0.6835&-0.40\\
Y&-8.79&273.13&05082002&15.08&0.96&3600&0.0454&0.6996&0.6952&-0.35\\
S4 R2 15&19.43&-150.88&05072002&15.21&0.95&3600&-0.0142&0.7138&0.7645&-0.25\\
S1 R2 3&125.25&71.33&05082002&15.24&0.97&3600&-0.0867&0.7252&0.7106&-0.25\\
S3 R5 4&-290.14&-264.51&05092002&15.25&0.98&3600&-0.1144&0.7983&0.8104&0.10\\
S4 R2 4&71.17&-131.34&05092002&15.30&0.94&3600&-0.0315&0.7085&0.7457&-0.30\\
S4 R4 22&-71.44&-276.28&05092002&15.31&0.93&3600&-0.0418&0.7017&0.7473&-0.35\\
S4 R4 27&-113.07&-277.44&05092002&15.42&0.93&1800&-0.1319&0.7170&0.7409&-0.25\\
S4 R4 26&-112.30&-267.91&05072002&15.47&0.93&5400&-0.1237&0.7603&0.7621&-0.10\\
O&55.48&-192.49&05072002&15.69&0.89&3600&-0.1223&0.7725&0.8202&0.00\\

\enddata
\tablenotetext{a}{Identifier from Cuffey (1965)}
\tablenotetext{b}{From Rey et al. (1998)}
\end{deluxetable*}

Our sample is comprised of 24 red giant stars in M53, selected from the photometric atlas of Cuffey (1965).  In order to have a data set as homogeneous and intercomparable as possible, all of the observations were made using the Kast double-beam spectograph on the Shane 3-meter telescope at Lick Observatory.  A mirror was used in place of a dichroic to direct all the stellar light to the blue side of the spectrograph, and the 452 l/mm grism and a slit width of 1.5$\arcsec$ produced a mean dispersion of 2.54 \AA\ /pixel and a mean resolution of 6.3 \AA.  Exposure times varied from 1800 to 5400 seconds, giving a signal-to-noise ratio per pixel at the 3883 \AA\ CN band between 20 and 60.

Photometry for the program stars was taken from the CCD study of Rey et al. (1998).  Cross-identification was done by taking the full online data set\footnote{Available at http://vizier.u-strasbg.fr/viz-bin/VizieR?-source=J/AJ/116/1775/}, plotting the locations of all stars brighter than 16th magnitude, and visually comparing that plot with the Digitized Sky Survey $R$-band image of M53.  Figure 1 is a color-magnitude diagram for the stars observed in M53, showing the reasonably uniform coverage of our data set over a range in absolute magnitude of $-0.5 < M_V < -2.5$. We adopt an apparent distance modulus of $(m-M)_V = 16.31$ from the online catalog of Harris (1996). All but four of the stars observed are consistent photometrically, spectroscopically, and in radial velocity with being first-ascent red giant branch stars.  M53 I, S1 R6 5, S3 R2 1, and S3 R2 15 (shown as crosses in Figure 1) are excluded from the sample because it is clear from their radial velocities and spectra that they are not cluster members.

Table 1 lists identifiers, positions, dates of observation, photometry, exposure times, three bandstrength indices, and measured carbon abundances for each star in our sample.  The X and Y offsets refer to the distance, in arcseconds, of each star from cluster center, and are given in the online data associated with Rey et al. (1998).  The results from the spectroscopy are listed as $S(3839)$, $S(CH)$, and $CH(4300)$.  The index $S(3839)$ is a measure of the 3883 \AA\ CN band, and $S(CH)$ and $CH(4300)$ (the later of which was introduced by Harbeck et al. 2003) both represent the depth of the 4300 \AA\ CH feature.  We define $S(CH)$ as 
\begin{displaymath}
S(CH) = -2.5 \log \frac{\int_{4280}^{4320} I_{\lambda}d\lambda}{\int_{4050}^{4100} I_{\lambda}d\lambda + \int_{4330}^{4350} I_{\lambda}d\lambda }
\end{displaymath}
where $I_{\lambda}$ is the number of counts per pixel at a given wavelength.  Values of $CH(4300)$ are given to facilitate comparisons with previous work.

The spectroscopic data were reduced using the XIDL suite of IDL programs developed by Jason Prochaska and Gabriel Prochter.\footnote{Available from http://www.ucolick.org/\char126 xavier/IDL/}  The Kast-specific programs were used to identify observation types, create a nightly HeHgCd composite arc spectrum and dome flat-field image, do flat-field corrections, trace, extract and wavelength-calibrate one-dimensional spectra, and combine multiple spectra of the same object.  The XIDL routines did not include an option for correcting spectra for atmospheric extinction; since some of our spectra were taken at an airmass of 2, a script was written to perform those corrections prior to the measurement of indices from the data.  The correction was based on the mean extinction curve given on the Lick Observatory web site.

We transformed the bandstrength indices onto a standard system defined by indices derived from HST spectra of flux standard stars taken from the STScI CALSPEC database described by Bohlin (1996).  These ''standard" index values were compared to indices measured from the extinction-corrected (but unfluxed) Kast spectra that were obtained on each night of one or more of the CALSPEC flux standards.  Figure 2 shows the differences (in the sense $S_{\rm{HST}}-S_{\rm{Kast}}$) for the indices $S(3839)$ and $S(CH)$ for each night of observation.  They are quite consistent from night to night, both within an observing run and between runs separated months apart.  These differences were applied as offsets to correct the Kast indices from the M53 spectra onto a standard system.

\begin{figure}[]
\includegraphics[width=\columnwidth]{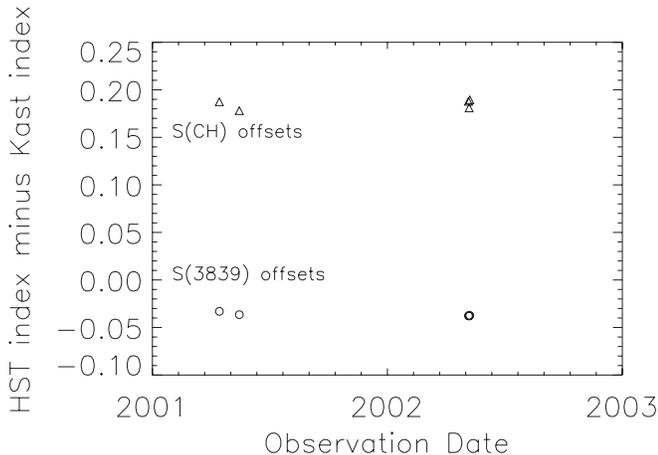}
\caption{
Nightly offsets between CH and CN bandstrength indices measured from extinction-corrected, unfluxed Kast spectra of flux standard stars and those measured from HST spectra (taken from the CALSPEC database at STScI, Bohlin 1996) of the same flux standards.  On 2001/04/15, the standard star observed was Kopff 27; on all other nights the standard star was Feige 34.}
\end{figure}

One-sigma errors in the index measurements were estimated by measuring $S(3839)$ and $S(CH)$ from individual spectra of stars observed more than once.  Typically each observation of a star consisted of two or three consecutive CCD exposures.  In cases where three observations were obtained of each star, we calculated $\sigma_{\rm {S(3839)}}$ as the standard deviation in the mean of the individual index values, and $\sigma_{\rm {S(CH)}}$ analogously.  In cases where only two observations were made per star, $\sigma_{\rm {S(3839)}}$ was calculated as $\frac{0.89 \times \Delta S(3839)}{\sqrt{2}}$, where $\Delta S(3839)$ refers to the difference between the two available values of the CN index, and $\sigma_{\rm {S(CH)}}$ analogously.  The resulting mean one-sigma index errors are $\langle \sigma_{\rm{S(3839)}} \rangle=0.02$ and $\langle \sigma_{\rm{S(CH)}} \rangle=0.007$.

Figure 3 shows $S(3839)$ versus $M_V$ for our data set (top panel) and NGC 6752 (bottom panel).  NGC 6752 has a metallicity of $-1.56$ (Harris 1996), and its $S(3839)$ distribution is typical of moderate-metallicity globular clusters, with two distinct sequences separated by a gap.  The data for NGC 6752 come from Norris et al. (1981).  The increase in bandstrength as temperature decreases is partly caused by more abundant molecule formation at lower temperature, but is made steeper by the changing shape of the stellar continuum.  The M53 data include the error bars calculated in the previous paragraph, and it is clear that the range in $S(3839)$ at a given $M_{V}$ is significant relative to $\sigma_{\rm{S(3839)}}$, indicating that the width of the CN variation is intrinsic to the stars rather than a result of measurement errors.  We identify two groups of M53 red giants in Figure 3 on the basis of their $S(3839)$ indices: a CN-weak group (plotted as open circles) and a CN-enhanced group (filled circles).

\begin{figure}[]
\plotone{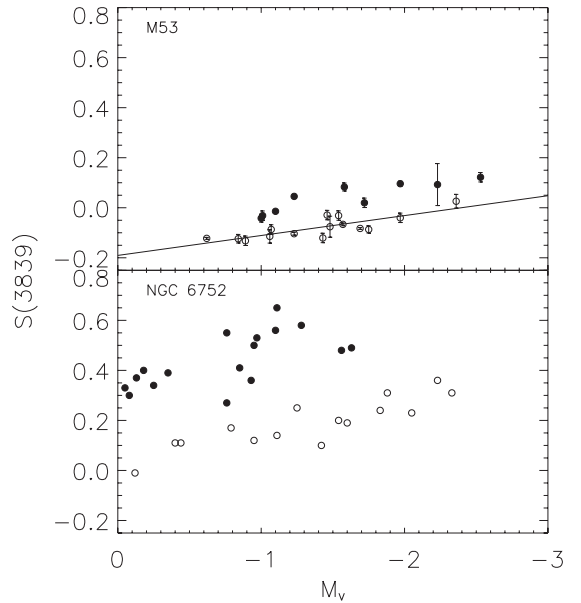}
\caption{
The 3883 \AA\ CN bandstrength index $S(3839)$ versus absolute $M_{V}$ magnitude for red giant stars in M53 (top panel) and NGC 6752 (bottom panel).  The upward linear trend exists because the shape of the stellar continuum and the abundance of CN molecules change with temperature.  Relatively CN-enhanced stars are shown as filled circles.  The straight line in the top panel is the baseline against which the quantity $\delta S(3839)$ is measured.}
\end{figure}

\section{The Models}
Conversion from the ($V$, $B-V$) plane to the ($T_{\rm eff}$, $\log g$) plane was accomplished by comparing the M53 fiducial sequence from Rey et al. (1998) to isochrones from the Victoria-Regina models (VandenBerg et al. 2006).  The 12 Gyr, [Fe/H]$=-1.84$, [$\alpha$/Fe]=+0.3 isochrone was the best match.  MARCS model atmospheres (Gustafsson et al. 1975) were created for 27 ($T_{\rm eff}$, $\log g$) pairs between $M_V = -0.5$ and $M_V = -2.6$ on the red giant branch of that isochrone.  The model parameters were set for consistency with the model spectra used in Briley \& Cohen (2001): [O/Fe]$=+0.2$, $^{12}{\rm C}/^{13}{\rm C}=10$, and $v_{turb}=2$ km/s.  The SSG spectrum synthesis code (Bell, Paltoglou, \& Trippico 1994 and references therein) was used to create model spectra for each of the 27 ($T_{\rm eff}$, $\log g$) points with [C/Fe] varying from $-1.0$ to $+0.5$ dex and [N/Fe] varying from $-0.5$ to $+1.0$ in steps of 0.1 dex.  Spectra were smoothed to a resolution of 5.4 \AA\ and a pixel spacing of 1.8 \AA\ to better match the data. 

\begin{figure}[]
\plotone{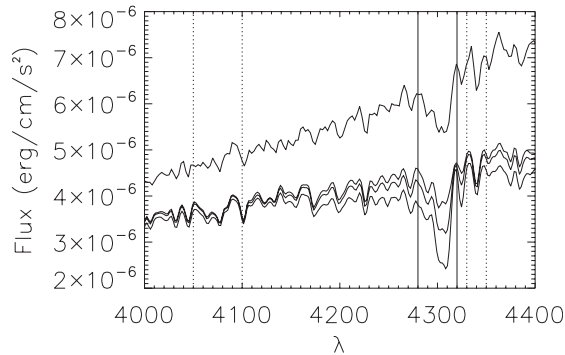}
\caption{
Synthetic spectra in the region of the 4300 \AA\ CH G band.  The spectrum of M53 star G is also shown, vertically offset from the models, for comparison.  Parameters are [C/Fe]$=(-1.0, -0.5, 0.0)$, [N/Fe]$=+0.6$, [O/Fe]$=+0.2$, $M_{V}=-1.5$, ${\rm ^{12}C/^{13}C=10}$, $v_{turb}=2$ km/s.  Vertical lines show the locations of the two sidebands (4050 \AA\ to 4100 \AA\ and 4330 \AA\ to 4350 \AA\ ) and the CH absorption bandpass (4280 \AA\ to 4320 \AA\ ) of the $S(CH)$ index.}
\end{figure}

\begin{figure}[]
\plotone{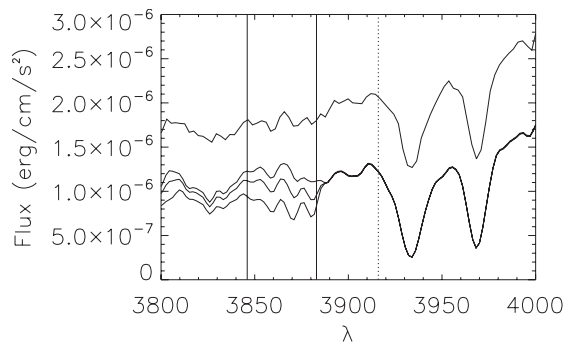}
\caption{
Synthetic spectra in the region of the 3883 \AA\ CN band.  The spectrum of M53 star G is also shown, vertically offset from the models, for comparison.  Parameters are [N/Fe]$=(0.0, 0.5, 1.0)$, [C/Fe]$=-0.3$, and all others the same as in Figure 4.  Vertical lines show the locations of the comparison sideband (3883 \AA\ to 3916 \AA\ ) and the CN absorption bandpass (3846 \AA\ to 3883 \AA\ ).}
\end{figure}

Figure 4 shows synthetic spectra which illustrate the strong dependence of the 4300 \AA\ G band on carbon abundance at $M_{V} = -1.5$ (with a fixed [N/Fe]).  The vertical dotted lines delineate the sidebands while solid lines define the science band for the $S(CH)$ index.  For comparison, the spectrum of M53 G is also shown, vertically offset from the synthetic spectra.  We chose the sidebands to lie close to the science band, but with as little CH absorption as possible.  The red sideband was necessarily narrow, as there is very little [C/Fe]-independent spectrum redward of 4320 \AA.  The blue sideband is farther from the science band than we would prefer, but there are complex CN molecular features near 4200 \AA.  Positioning of the blue sideband (4050 \AA\ to 4100 \AA\ ) is a compromise between band width, location, and [C/Fe] dependence.

\begin{figure}[]
\plotone{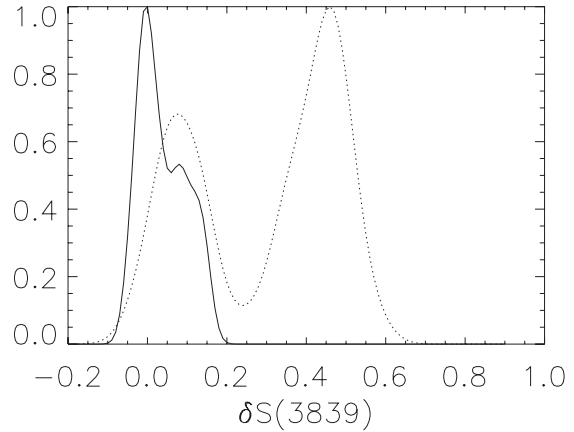}
\caption{
Generalized histogram of $\delta S(3839)$ for M53 red giant stars (solid line).  The dotted line is the analogous curve for NGC 6752 red giants (data from Norris et al. 1981).  Each histogram is normalized to a maximum of unity.  A generalized histogram is the sum of unit Gaussian curves centered at each data point, with the dispersion $\sigma$ of each Gaussian equal to the error on the points (which was taken to be 0.02 for M53 and 0.045 for NGC 6752).}
\end{figure}

Figure 5 shows the dependence of the 3883 \AA\ CN band on nitrogen abundance at $M_{V}=-1.5$ (with a fixed [C/Fe]), and the wavelength bands for the $S(3839)$ index are marked with vertical lines.  Only the red edge of the comparison band is shown as a dotted line because its blue edge coincides with the red limit of the science band.  For comparison, the spectrum of M53 G is also plotted, vertically offset from the three synthetic spectra.  The red sideband for $S(3839)$ is far easier to place than either of the sidebands for $S(CH)$, as there is a wide region redward of the 3883 \AA\ bandhead that is insensitive to [N/Fe].  The strong \ion{Ca}{2} K line is avoided.  It is impractical to place a comparison sideband to the blue of the 3883 \AA\ CN feature, since the intrinsic flux of a red giant drops precipitously into the UV.  There is also a forest of metal absorption lines and hydrogen Balmer lines in the UV, and furthermore it is problematic to find a wide spectral region to the blue that is reasonably independent of [C/Fe] and [N/Fe].  The single sideband is partially responsible for the strong temperature trend in $S(3839)$, and can also cause significant instrument-to-instrument offsets.  For these reasons Norris et al. (1981) introduced the quantity $\delta S(3839)$, defined as the measured $S(3839)$ minus a baseline value that is determined from a linear fit to the lower envelope of the CN-weak stars in the ($S(3839)$,$M_{V}$) plane.  The baseline adopted for the M53 red giants is $S_{0}=-0.191-0.080M_{V}$.

\section{Analysis}
We are interested in comparing the CN and CH index behavior in M53 to that in globular clusters with more moderate metallicity, and use two separate methods for making this comparison.  In an index analysis, we compare the behavior of the $S(3839)$ and $S(CH)$ indices in M53 with the variations observed in NGC 6752, a typical example of a CN-bimodal globular cluster.  In an abundance analysis, we determine [C/Fe] abundances for individual M53 red giants by comparing data to isoabundance loci in the $S(CH)$ versus $M_{V}$ plane.

\subsection{Index analysis}
NGC 6752 is an excellent example of bimodal CN bandstrength and anticorrelated [C/Fe] and [N/Fe] behavior.  Figure 6 shows a generalized histogram of $\delta S(3839)$ for M53 red giants (solid line) with the analogous NGC 6752 data from Norris et al. (1981) overplotted as a dotted line.  The solid histogram is a sum of Gaussian curves with a $\sigma$ of 0.02, the standard error on $S(3839)$ for the Kast M53 spectra, while the Gaussian $\sigma$ used by Norris et al. (1981) is 0.045.  The distribution of CN bandstrength in M53 is clearly not well-described by a single symmetric Gaussian curve; indeed, a two-sided Kolmogorov-Smirnov test comparing $\delta S(3839)$ for stars identified as CN-weak and CN-enhanced in Figure 3 returns a probability of $2.67\times10^{-5}$ that the two groups were drawn from the same underlying distribution.  However, there is a clear difference between the M53 CN distribution and that of NGC 6752: the latter has a large separation between two clearly distinct CN groups, and also has a more well-populated CN-strong group than the former.  

\begin{figure}[]
\plotone{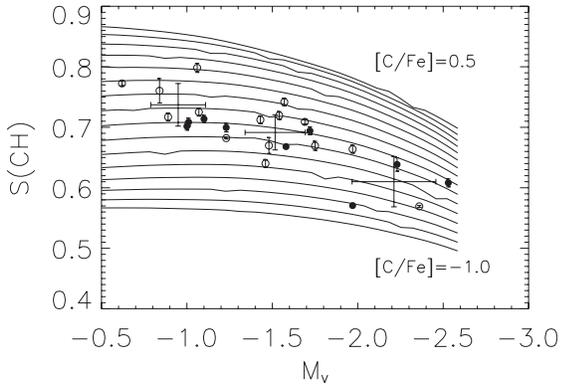}
\caption{
The 4300 \AA\ CH bandstrength index $S(CH)$ versus $M_{V}$, with CN-enhanced stars (identified in Figure 3) plotted as filled circles.  $S(CH)$ declines with decreasing $M_{V}$, and at magnitudes of $M_{V}>-1.4$ the CN-enhanced stars are slightly offset as a group in $S(CH)$ from CN-weak stars.  The error bars on individual data points correspond to the measurement errors $\sigma_{\rm S(CH)}$ as described in $\S$2.  The large crosses represent the mean values and standard deviations for the data in three luminosity bins: $-0.5 \geq M_{V} \geq -1.2$, $-1.2 \ge M_{V} \geq -1.9$, and $-1.9 \ge M_{V} \geq -2.6$.  Isoabundance contours from synthetic spectra are shown as solid lines.  Each curve has the parameters [N/Fe]$=+0.6$, [O/Fe]$=+0.2$, $-2.6 \le M_{V} \le -0.5$, ${\rm ^{12}C/^{13}C}=10$, and ${v_{turb}}=2$ km/s. [C/Fe] varies from $-1.0$ dex (bottom line) to $+0.5$ dex (top line), with a spacing of 0.1 dex. It can be seen from the models that $S(CH)$ is sensitive to temperature in stars brighter than $M_{V}\simeq -1.4$. Mean $S(CH)$ values for the three luminosity bins decline faster than can be attributed to temperature effects alone, indicating that carbon is progressively depleted as stars ascend the giant branch.}
\end{figure}

Figure 7 shows $S(CH)$ versus $M_{V}$ for our M53 data, with stars identified as CN-enhanced (see Figure 3) plotted as filled circles.  There is a general decline in $S(CH)$ with $M_{V}$, and there is also an anticorrelation between CN bandstrength and $S(CH)$ among the fainter stars ($M_{V} > -1.4$).  In the later case, the CN-enhanced giants have values of $S(CH)$ at the low end of those found among the CN-weak giants, although there are only four CN-enhanced giants in our sample with $M_{V} > -1.4$. Among brighter giants with $M_{V} < -1.6$ there is no such trend.  The large crosses represent the mean values and standard deviations of the M53 data binned into three luminosity groups: $ -0.5 \ge M_{V}\geq -1.2$, $-1.2 \ge M_{V} \geq -1.9$, and $-1.9 \ge M_{V} \geq -2.6$.  We have overplotted model isoabundance lines for [C/Fe] varying from $-1.0$ dex (bottom) to $+0.5$ dex (top).  All of the models plotted here were computed for the parameters 
[N/Fe]$=+0.6$, [O/Fe]$=+0.2$, ${\rm ^{12}C/^{13}C}=10$, and $v_{turb}=2$ km/s.  The comparison between the data and the isoabundance curves reveals two main features.  First, at a fixed [C/Fe], the model curves show a clear temperature effect on $S(CH)$ for stars brighter than $M_{V}=-1.4$ ($T_{\rm eff} \leq 4500$ K).  Second, the decline in $S(CH)$ with rising luminosity in the data is steeper than the temperature effect alone, and this implies that there is a decline in [C/Fe] with decreasing $M_{V}$ among the red giants in M53.

Figure 8 shows $\delta S(3839)$ versus $S(CH)$ for our M53 data.  The left plot shows the full data set, and the right panel shows only the stars fainter than $M_{V}=-1.4$.  There is no clear anticorrelation in the left panel, whereas the right panel shows that among stars with $M_{V} > -1.4$
those with enhanced CN bands have relatively low $S(CH)$ indices, as noted also from Figure 7.  The lack of a more prominent anticorrelation between CN and CH in the full data set may be partly due to a reduced sensitivity of the index $S(3839)$ to nitrogen abundance at the low metallicity of M53.  This point is illustrated by Figure 9, which shows the $S(3839)$ measurements for M53 versus $M_{V}$, with model isoabundance curves for [C/Fe]$=-0.3$ dex and [N/Fe] ranging from $-0.5$ to $+1.0$ dex overplotted.  Since (as we show below) not all of the stars have [C/Fe]$=-0.3$ dex, the distribution of data points in this plot is not necessarily indicative of true nitrogen abundance.  At the location of our faintest star in the ($M_{V}, S(3839)$) plane, changing [N/Fe] by 0.1 dex changes $S(3839)$ by 0.02.  Since a factor of two change in nitrogen abundance only produces a few-sigma change in $S(3839)$, its usefulness as an indicator of [N/Fe] is limited at the metallicity of M53.

\begin{figure}[]
\plotone{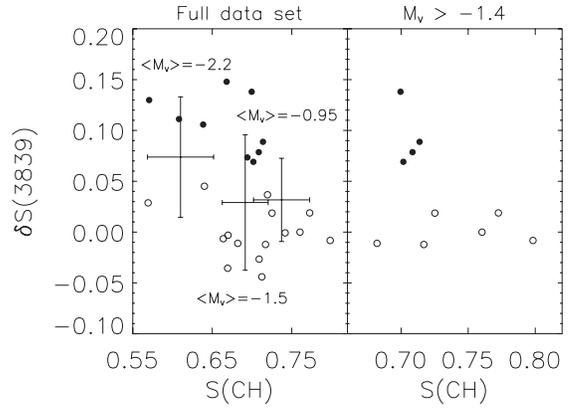}
\caption{
Left panel: The 3883 \AA\ CN bandstrength index $\delta S(3839)$ versus $S(CH)$ for all red giants in the sample, with stars identified as CN-enhanced in Figure 3 shown as filled circles.  The three crosses represent the mean and standard deviation in the same luminosity bins introduced in Figure 7.  There is evidence for some deep mixing: stars in the brightest bin, on average, have smaller $S(CH)$ and higher $\delta S(3839)$ than stars in the two fainter bins.  Right panel: $\delta S(3839)$ versus $S(CH)$ for red giants with $M_{V} \ge -1.4$.  }
\end{figure}

Since $S(3839)$ has a diminished sensitivity to [N/Fe] at the metallicity of M53, we do not attempt to use it to measure [N/Fe] for individual stars.  However, we can comment on the range of [N/Fe] in our sample, and compare it to the range reported by Norris et al. (1981) between red giants in NGC 6752.  Figure 10 shows $S(3839)$ versus $M_{V}$ for the Kast M53 data, with two sets of models overplotted: the lower line is intended to represent a typical CN-weak star with [C/Fe]$=-0.3$ and [N/Fe]$=0$.  This abundance combination was chosen to produce a model locus that follows the lower envelope of the M53 stars in Figure 10.  The upper line represents typical CN-strong stars with [C/Fe]$=-0.6$ and [N/Fe]$=0.9$.  The abundance differences between the CN-weak and CN-strong loci were chosen to match those reported in NCG 6752 by Norris et al. (1981).  They measured the abundance differences between a pair of CN-weak and CN-strong giants in NGC 6752 with $M_{V}=-1.3$ as $\Delta$[N/Fe]=0.9, $\Delta$[C/Fe]$=-0.3$.  Similar differences were found by Da Costa \& Cottrell (1980) for another pair of giants in NGC 6752 that are one magnitude fainter.  The two isoabundance curves in Figure 10 bound most of the data points for M53, meaning that while some pairs of stars in this cluster may have carbon and nitrogen differences as large as those in NGC 6752, the majority of giants in M53 have a smaller dispersion.  

\begin{figure}[]
\plotone{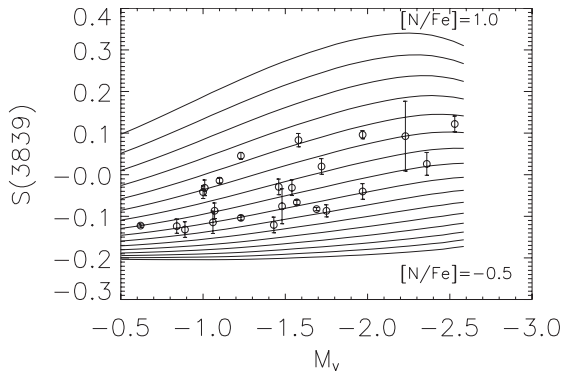}
\caption{
The 3883 \AA\ CN bandstrength index $S(3839)$ versus $M_{V}$ for stars in M53.  Error bars are the measurement errors $\sigma_{\rm S(3839)}$ as described in $\S$2.  Isoabundance curves are shown as solid lines; their parameters are [C/Fe]$=-0.3$, $-0.5 \le$ [N/Fe] $\le 1.0$, and all others the same as in Figure 7.  This plot is meant to illustrate the dependence of $S(3839)$ on nitrogen abundance at low metallicity: according to the models, there is only a small change in $S(3839)$ for a given change in [N/Fe] particularly in the region of this figure occupied by the CN-weakest giants.  Since the model isoabundance contours plotted here have [C/Fe] set to $-0.3$ dex, and not all of the M53 red giants have such a carbon abundance, this plot is not meant to be indicative of [N/Fe] abundance for any star in particular.}
\end{figure}

\subsection{Carbon abundances}
To calculate carbon abundances for the stars in our M53 sample, we compare the $S(CH)$ values from the Kast spectra to those from the model spectra.  Figure 7 presents the data used for determining [C/Fe], and shows $S(CH)$ versus $M_{V}$ for both data (circles) and models (solid lines).  We interpolate linearly between the model isoabundance contours to calculate [C/Fe] for each star.  The vertical error bars on the individual data points represent the observational uncertainty in $S(CH)$.  The mean value of that error is $\sigma_{\rm S(CH)}=0.007$, which is equivalent to an error in [C/Fe] of roughly 0.05 dex.  The derived values of [C/Fe] are accordingly rounded to the nearest multiple of 0.05 dex.  Figure 11 shows [C/Fe] versus $M_{V}$ for the CN-enhanced and CN-weak stars (as identified in Figure 3), and it shows a clearly declining trend, at least among the CN-weak giants, as is suggested by Figure 7.  The carbon depletion rate in the full M53 sample is 0.22 dex per V magnitude, which is similar to the rate measured in red giants in the halo field by Smith \& Martell (2003).  Studies of red giants in other globular clusters also show abundance trends that indicate ongoing deep mixing (see, e.g., Carbon et al. 1982 and Trefzger et al. 1983).   The theoretical study of Denissenkov \& VandenBerg (2003), which models deep mixing as a diffusive process, finds that deep mixing rates among low-mass red giants ought to be fairly independent of mass and metallicity, which may explain why abundance trends are so similar among so many different groups of red giants.

However, there is the possibility that the measured carbon abundances are affected by a systematic error resulting from initial model assumptions.  Specifically, when the synthetic spectra were created, it was assumed that all stars had the same oxygen abundance regardless of luminosity.  By not allowing for a decline in [O/Fe] with rising luminosity, our models pertain to a case in which deep mixing has not affected oxygen.  

Free oxygen in cool stellar atmospheres ($T_{\rm eff} \leq 4500$ K) participates in molecular equilibrium by forming CO, which consumes carbon atoms that could otherwise form CH molecules.  With a fixed carbon abundance, therefore, an oxygen-enhanced star will have weaker CH bands than an oxygen-depleted star.  If [O/Fe] starts at a value of +0.2 dex, but oxygen is depleted as stars evolve along the giant branch, due to some mixing process that our models do not take into account, then too large a carbon abundance will be inferred for the most evolved stars, since we will have assumed too large an oxygen abundance for them.  This will act to make the decline in Figure 11 artificially shallow, and the true decrease in [C/Fe] with brightening $M_{V}$ would be steeper than indicated.

\begin{figure}[]
\plotone{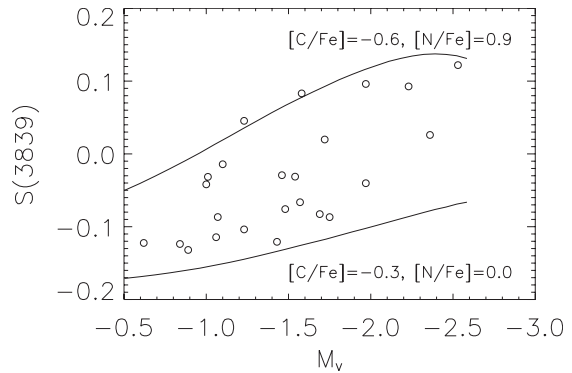}
\caption{
The 3883 \AA\ CN bandstrength index $S(3839)$ versus $M_{V}$ for our sample of M53 red giants, with two sets of models overplotted: the lower line is typical of a CN-weak giant with [C/Fe] set to $-0.3$ dex and [N/Fe] to $0.0$ dex, and the upper line is typical of a CN-strong star, with [C/Fe]$=-0.6$ dex and [N/Fe]$=+0.9$ dex.  The anticorrelated abundance difference between the two model tracks is taken from the Norris et al. (1981) and Da Costa \& Cottrell (1980) studies of NGC 6752, a typical CN-bimodal globular cluster.  Some of the M53 stars fall on or near the model lines, but most lie between, indicating that they have intermediate carbon and nitrogen abundances.}
\end{figure}

Briley et al. (1990) consider the effect of varying [O/Fe] from $-0.6$ to $+0.4$ dex on their [C/Fe] measurements for the low-metallicity cluster M55.  Their Table 7 shows that [C/Fe] determined from CH bandstrength can change by as much as 0.5 dex for a 1.0 dex change in assumed [O/Fe], with the effect being largest in the coolest stars.  Smith \& Briley (2006) collected abundance data from the literature for the moderate-metallicity globular cluster M13.  The data show a considerable spread in [O/Fe] at a given $M_{V}$, and a decline of roughly 1 dex in the minimum [O/Fe] over the $M_{V}$ range covered in our M53 sample.  If oxygen behaves in an analogous way in M53, our use of constant-[O/Fe] models would cause systematic errors in the [C/Fe] difference inferred between the faintest stars in our sample and those brighter giants that have been diminished in oxygen.  However, as noted in the previous paragraph, such an effect would not lead us to infer a spurious decline in carbon abundance along the red giant branch, but would act in the opposite sense to partially mask any carbon decrease that is present. Thus, the evidence of Figure 11 that evolutionary changes do occur in the surface carbon abundances of the M53 red giants would remain valid, although the size of the effect may be underestimated in the present work.

\begin{figure}
\includegraphics*[width=\columnwidth,keepaspectratio=true]{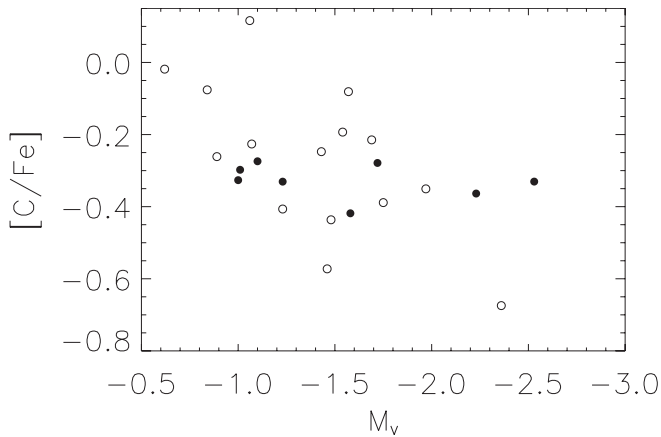}
\figcaption{
Carbon abundance versus \protect$M_{V}$ for red giants in M53, with CN-enhanced stars (identified in Figure 3) plotted as filled circles.  The decline in [C/Fe] with increasing luminosity expected from Figure 7 is seen here.  As discussed in the text, it is possible that [C/Fe] is overestimated for the CN-strong stars brighter than \protect$M_{V}\sim-1.4$ if overestimates of [O/Fe] are used in the models.  }
\end{figure}

The hottest stars in our M53 sample should be the least affected by [O/Fe] variations because they are unlikely to have significant abundances of CO in their atmospheres.  Suntzeff (1981) showed that relatively little CO forms above 4500 K in red giant photospheres at the metallicity of M13.  According to the Victoria isochrone we have adopted for M53, a temperature of 4500 K corresponds to an absolute magnitude of $M_{V}=-1.4$.  It is among stars fainter than this that the evidence for a CN-CH band anticorrelation in Figure 8 is greatest.  If the atmospheres of CN-enhanced stars have been subjected to greater CNO-processing than those of the CN-weak stars, they will have lower [O/Fe] and [C/Fe] abundances but higher [N/Fe] than CN-weak stars.  This will produce a CN-CH band anticorrelation in stars hot enough to have little or no CO molecular formation in the atmosphere. However, in cooler CN-enhanced giants there would be less CO formation than in the CN-weak giants, which may partially counteract a CN-CH anticorrelation by leaving relatively more carbon available for CH formation.  Figure 8, seen in this light, shows a CN-CH anticorrelation in the hotter stars that may be erased as CO formation sets in more strongly in the CN-weak stars.

\section {Discussion}

The generalized histogram of $\delta S(3839)$ shows that M53 has a distribution of CN bandstrength whose width is greater than our 3-$\sigma$ errors.  We infer that M53 has a population of CN-enhanced giants, and not just CN-weak stars.  However, there are two physical factors that erode the appearance of CN bimodality at low metallicity:  first, in a cluster with a low [Fe/H], the low carbon and nitrogen abundances can sustain only limited molecular CN formation.  Second, deep mixing scenarios predict that mixing efficiency should be high at low metallicity (e.g., Sweigart \& Mengel 1979; Charbonnel 1995), meaning that evolved stars in a low-metallicity cluster like M53 will be depleted in carbon and enhanced in nitrogen from the dredge-up of C$\rightarrow$N and possibly O$\rightarrow$N-processed material.  If the mixing continues to the point where carbon is less abundant than nitrogen, the abundance of CN molecules can decline with rising luminosity (see, e.g., the discussion of extremely C-depleted stars in Cohen et al. 2002 and Langer et al. 1985).

The relatively small range in carbon and nitrogen abundance variation observed in M53 may also be an indication that the polluting material was not processed through the full CNO cycle.  If this is the case, then M53 should not show an O-Na anticorrelation, because neither the O$\rightarrow$N processing nor the NeNa-cycle processing that occurs at the same temperature will have taken place.  The data collected for this study do not allow for an investigation into this question, but a followup study at higher resolution could be quite interesting.  O-Na anticorrelations have been found in red giants in every globular cluster where they have been sought, meaning that their absence in M53 would be notable.

In summary, we find a progressive decrease in [C/Fe] with brightening $M_{V}$ among red giants in M53, indicating that deep mixing is occurring.  The [C/Fe] and $S(3839)$ measurements further reveal that some stars have carbon depletions and nitrogen enhancements (relative to the cluster CN-weak stars) as large as those observed in the CN-strong giants of the classically CN-bimodal globular cluster NGC 6752 (although we do not see a strongly bimodal CN distribution in M53), indicating that whatever process is responsible for primordial abundance variations in NGC 6752 (as described in Caretta et al. 2005 and Suntzeff \& Smith 1991) is also operating at the lower metallicity of M53 ([Fe/H]$\sim -1.8$).  We conclude that the rarity of CN-strong stars in other clusters with [Fe/H] $\sim-1.8$ or lower, such as M55, is more likely an effect of the overall low metallicity causing diminished CN molecule formation and the reduced sensitivity of the index $S(3839)$ at low metallicity, and is not due to some unexplained lack of abundance inhomogeneity\footnote{However, it is certainly the case that the fraction of CN-strong giants varies among globular clusters, as discussed in, e.g., Norris 1987.  The small number of M53 giants that lie near the upper isoabundance line in Figure 10 may indicate that M53 has a lower fraction of CN-enhanced stars than NGC 6752.}.  A direct test of this hypothesis would be to obtain spectroscopy of the $\lambda$3360 NH band for red giants in M53 and M55.  In the case of M92, which is of even lower metallicity, it would seem that the formation of CN-strong giants is inhibited by the low overall metallicity of [Fe/H] $\sim -2.3$ even though large [C/Fe] and [N/Fe] differences, as well as large [N/Fe] enhancements, were found among the red giants by Carbon et al. (1982).

\acknowledgements
The authors acknowledge the support of NSF grant AST 04-069888.  We thank the Lick Observatory Time Allocation Committee for providing observing time for this project, and the support astronomers at Mount Hamilton for their assistance at the 3-m telescope.  SLM also thanks the ARCS Foundation for its support.

\end{document}